\documentclass{PoS}

 \title{Precision test of the gauge/gravity duality in two-dimensional  N=(8,8) SYM}

 \ShortTitle{Precision test of the gauge/gravity duality in two-dimensional N=(8,8) SYM}

 \author{\speaker{Daisuke Kadoh}%
         \thanks{This work was done in collaboration with Eric Gigu\`ere.}
         \\
        Department of Physics at Hiyoshi, and Research and Education Center for Natural Sciences,
Keio University, Hiyoshi 4-1-1, Yokohama, Kanagawa 223-8521, Japan\\
        E-mail: \email{kadoh@keio.jp}}

 \abstract
 {The $\epsilon - p$ is calculated from lattice simulations of two dimensional ${\cal N}=(8,8)$ $SU(N)$ SYM to test the gauge gravity duality. We employ the Sugino action with keeping two of sixteen supercharges exactly on the lattice. The thermodynamics of this gauge theory is described by the black 1-branes at low temperature. The internal energy density $\epsilon$ and the pressure $p$ can be analytically estimated in the gravitational theory. The lattice results in $N=12$ show that $\epsilon - p$ of the thermal SYM reproduces the prediction of the gravitational theory for the dimensionless temperature $T_{\rm eff}<0.4$. This clearly indicates that the duality is valid in this system. 
 }

 \FullConference{34th annual International Symposium on Lattice Field Theory\\
                  24-30 July 2016\\
                  University of Southampton, UK}

\begin{document}

 \section{Introduction}
 The gauge/gravity duality is a theoretical conjecture that is originally motivated by AdS/CFT \cite{Maldacena:1997re}, 
 and are widely accepted and  used to understand various phenomena 
in QCD, black holes, or even in condensed matter physics.
Such applications rely on the validity of the conjecture. 
We have to show that it is true in order to establish the applications from fundamental perspective and to develop further applications.
The numerical verification of the duality has been done for SYM with sixteen supercharges in several dimensions,  
\cite{Hanada_Nishimura}-\cite{Berkowitz:2016tyy},
and  some of them use the lattice formulations with a few exact supersymmetries, \cite{Kaplan:2005ta, Sugino:2004uv}. 
In this paper, we test the duality in two-dimensional $SU(N)$ supersymmetric Yang-Mills theory (SYM) with sixteen supercharges 
from the lattice simulations with the Sugino action at finite temperature. 
 
The description of the duality in this case is based on $N$ D1-branes of type-IIB superstring theory, 
which corresponds to the black 1-branes (black strings).
In the gravitation theory, the black string  thermodynamics is analytically treated  at low temperature (the large 't Hooft coupling limit) in the large N limit, and the free energy can be given as a function of temperature. On the other hand, in this limit, the gauge theory becomes a strongly coupled one of which non-perturbative methods are needed to clarify the physics.
We use the lattice theory, particularly Sugino's lattice formulation of  ${\cal N}=(8,8)$ SYM with keeping two exact supersymmetries, in order to estimate thermodynamical quantities in the gauge theory beyond the perturbation theory.

This paper is organized as follows: The continuum  ${\cal N}=(8,8)$ SYM is defined and the duality is introduced in section 2. 
We will review Sugino's formulation of  ${\cal N}=(8,8)$ SYM in section 3. 
Then, we will see the lattice results of the difference between the internal energy density and the pressure and compare them with the theoretical predictions of the gravitational theory to verify the duality in section 4. 
The section 5 is devoted to a conclusion and discussion.

 \section{SYM and the duality}
 We start with defining the supersymmetric Yang-Mills theory with sixteen supercharges in two dimensions. 
 This theory contains  the gauge fields $A_\mu\, (\mu=0,1)$
 and the eight  scalar fields $X_i\, (i=2,3,\cdots,9)$
  which are supersymmetric partners of the fermions $\psi_\alpha\, (\alpha=1,2.\cdots,16)$. We assume that the gauge group is $SU(N)$ 
  and all fields are expanded by the group generators $T^a (a=1,\cdots N^2-1)$
 such as $A_\mu(t,x) =A_\mu^a(t,x) T^a$ where the spacetime coordinate is represented by $(t,x)$. The spacetime is compactified to a torus,  
 and $\beta$ and $L$ are the sizes of time and space directions, respectively.

 The euclidean action is
 \begin{eqnarray}
&& 
       S=\frac{N}{\lambda}\int{}{\rm d}^2x\ {\rm tr} 
             \left\{ \frac{1}{4}F^2_{\mu\nu} 
                    + \frac{1}{2}(D_\mu X_i)^2 - \frac{1}{4}[X_i,X_j]^2  
             \right.  \nonumber \\
&& 
      \hspace{3cm} \left.
          +\frac{1}{2} \psi_\alpha  D_0 \psi_\alpha 
          -\frac{i}{2} \psi_\alpha (\gamma_{1})_{\alpha \beta} D_1 \psi_\beta 
        + \frac{1}{2}\psi_\alpha (\gamma_i)_{\alpha \beta} [X_i, \psi_\beta] \right\},
        \label{cont_action}
\end{eqnarray}
 where the matrices $\gamma_1, \cdots,\gamma_9$ satisfy the nine-dimensional Clifford algebra, $\{\gamma_i,\gamma_j\}=2\delta_{ij}$.
 The field tensor and the covariant derivative are defined as
  $F_{01} = \partial_0 A_1 - \partial_1 A_0 +i[A_0,A_1]$ and $D_\mu \varphi = \partial_\mu \varphi + i[A_\mu, \varphi]$, respectively.
 If the periodic boundary condition  is imposed on the fields for both space and time directions, the action (\ref{cont_action}) possesses supersymmetry with sixteen supercharges 
 since gauge, Yukawa and  $\phi^4$-interactions appear in a specific relative weights with single 't Hooft coupling constant $\lambda$.
   
 In finite temperature SYM theory, 
 the boundary condition for the time direction on the bosonic fields $B(=A_\mu,X_i)$ and that on the fermionic fields $\psi$ are different:
 \begin{eqnarray}
 B(t+\beta, x) = B(t, x),\quad \psi(t+\beta, x) = -\psi(t, x),
 \label{boundary_condition}
 \end{eqnarray}
 with inverse temperature $\beta=1/T$, while the same periodic boundary conditions with period $L$ are imposed for the space direction.
Then, supersymmetry is broken by the thermal effects, and it is naively expected to be restored in the zero temperature limit.

The system of $N$ coincident D1-branes is the gravity dual of SYM with sixteen supercharges in two dimensions. 
 The black 1-branes (black strings) solution is given by the following metric, 
  \begin{eqnarray}
 ds^2= H^{1/4}(r) \Big(
 H^{-1}(r)[f(r)dt^2+dx^2]
 + f^{-1}(r) dr^2 
 + r^2 d\Omega_7^2
 \Big),
  \end{eqnarray}
  with
   \begin{eqnarray}
   && H(r)=1+ {\rm sinh}^2 \gamma \frac{r_0^6}{r^6},\\
   && f(r) = 1- \frac{r_0^6}{r^6},
 \end{eqnarray}
 where $r_0$ is the radius of event horizon of the black 1-branes, and $\gamma$ is a variable corresponding to the charge of the branes. (see \cite{Klebanov:1996un} for more detailed definitions).
  As shown in \cite{Klebanov:1996un}, 
 near zero temperature, the mass (internal energy) and the entropy of the black 1-branes can be written as the functions of temperature (in the $\lambda=1$ unit):
  \begin{eqnarray}
 E = \frac{2}{3} c_0 L N^2 T^3, 
 \qquad S = c_0 L N^2 T^2,
 \end{eqnarray}
 where the numerical constant $c_0=2^4 \pi^{5/2}/3^3(=10.366...)$.
 The pressure $p$ is obtained through $E=TS-pV$ where the system volume $V=L$ in the present case.
The difference between the energy density and the pressure is thus given by
\begin{eqnarray}
\epsilon - p = \frac{c_0}{3} N^2 T^3.
\label{e_minus_p} 
\end{eqnarray}
As seen in the section 4, this combination plays an essential role to test the duality in this paper.

 \section{Sugino's formulation}
 We define the Sugino action for  ${\cal N}=(8,8)$ SYM  on a two-dimensional lattice \cite{Sugino:2004uv} with the lattice spacing $a$.
The lattice sites are labeled by integers: 
$t=1, \cdots, N_t, \, x=1,\cdots, N_x$ in the lattice spacing unit, where $N_t$ and $N_x$ are the temporal and spatial lattice sizes, respectively.
The gauge fields are expressed by gauge group-valued link fields $U_\mu(x)$ defined on the links
and the other fields are defined on the sites. Hereafter we set $a=1$ for simplicity.
 
  The Sugino lattice action of $N=(8,8)$ SYM is given by
 \begin{eqnarray}
&& S = Q_+Q_-\frac{N}{2\lambda_0}\sum_{t,x} \ {\rm tr}  
   \left\{  - 4i B_i F_{i3}^+ 
             - \frac{2}{3} \epsilon_{ijk} B_i B_j B_k
%
           - \psi_{+\mu}\psi_{-\mu}
           - \chi_{+i}\chi_{-i}
           - \frac{1}{4}\eta_+\eta_- \right\},
           \label{lat_action}
\end{eqnarray} 
with
\begin{eqnarray}
&& F_{03}^+ = \frac{1}{2}(\nabla^+_0   A_3 + \nabla^+_1   A_2 ), \label{Fp01} \\
&& F_{13}^+ = \frac{1}{2}(\nabla^-_1 A_3 - \nabla^-_0 A_2 ), \label{Fp02} \\
&& F_{23}^+ = \frac{1}{2}(i[A_2 ,A_3 ] + F_{01}),
\label{Fp03}
\end{eqnarray}
where 
\begin{eqnarray}
&&\nabla_\mu^+ \varphi (x) = U_\mu(x)\varphi (x+\hat\mu) U^{-1}_\mu(x) -  \varphi (x),\\
&&\nabla_\mu^+ \varphi (x) =   \varphi (x)- U_\mu(x-\hat\mu)^{-1}\varphi (x-\hat\mu) U_\mu(x-\hat\mu),
\end{eqnarray}
are the forward and the backward covariant difference operators, respectively, and the lattice field tensor $F_{01}$ is defined as
\begin{eqnarray}
&& F_{01}(x)=-\frac{i}{2}\left(P_{01}(x) - P_{01}^\dagger(x) - \frac{1}{N} {\rm tr}(P_{01}(x) - P_{01}^\dagger(x))\right),
\label{F01}
\\
&& P_{01}(x)=U_0(x)U_1(x+\hat{0})U_0^\dagger(x+\hat{1})U_1^\dagger(x).
\label{P01}
\end{eqnarray}
The lattice (infinitesimal) gauge transformations,
\begin{eqnarray}
\delta_\omega U_\mu (x) =i\omega(x) U_\mu (x) - i U_\mu (x) \omega(x+\hat\mu), \qquad \ \ 
 \delta \varphi(x) = -i [\varphi(x), \omega(x)]
\label{lat_gauge_transf}
\end{eqnarray}
make the action (\ref{lat_action}) invariant since all fields in the integrand in it transform in the gauge covariant way.
In addition,  all fields satisfy the boundary condition (\ref{boundary_condition}) and supersymmetry is explicitly broken.        

Seemingly, this lattice action (\ref{lat_action}) does not look like
the continuum action (\ref{cont_action}) in the continuum limit:  
It is written in a  $Q_\pm$-exact form and the integrand is a function of $B_i,\psi_{\pm \mu}, \eta_\pm,\chi_{\pm i}$ 
which are given by linear combinations of the original variables, $X_i, \psi_\alpha$.  
The $Q_\pm$ are two of sixteen supersymmetry transformations which satisfy
$Q_\pm^2 = i \delta_{\phi_\pm}$ and $\{Q_+,Q_-\} = -i\delta_C$ where seven auxiliary fields are introduced to realize the off-shell nature. 
The new variables are useful to express $Q_\pm$-transformations as simple forms.  
The change of variables and the $Q_\pm$-transformations are explicitly given in \cite{Giguere:2015cga}. 
After performing $Q_\pm$ in (\ref{lat_action})  and getting back to the original variables, 
 (\ref{lat_action}) coincides with (\ref{cont_action}) in the continuum limit.

The lattice spacing effect of supersymmetry breaking vanishes in the classical continuum limit, 
however some  SUSY breaking operators can be radiatively generated and are relevant as the quantum theory in general.  
In the perturbation theory, two exact supersymmetires $Q_\pm$ forbid the SUSY breaking operators and sixteen supersymmetries are restored even at the quantum level. 
In \cite{Giguere:2015cga}, the numerical results of the SUSY Ward-Takahashi-relation suggest us that the SUSY restoration does occur 
beyond the perturbation theory thanks to the two exact supersymmetries $Q_\pm$.
 
 The simple $F_{01}$ (\ref{F01}) which is used to define (\ref{lat_action}) causes the problem of the extra vacua as mentioned in  \cite{Sugino:2003yb}. There are a few possible choices for $F_{01}$ to avoid the problem: an admissibility-type field tensor\cite{Sugino:2003yb} or a tangent-type field tensor\cite{Matsuura:2014pua}.  The lattice action with these field tensors are theoretically well-supported 
 in the sense that the extra vacua are completely forbidden. However simulation codes become rather complicated ones for them, and we chose the simple one (\ref{F01}).
 As shown in \cite{Giguere:2015cga}, 
 this choice does not cause the problem in  practice  in the numerical simulations 
 with the Hybrid Monte-Carlo method since the extra vacuum configurations are not observed in many parameters we are interested in.

 \section{Lattice results}
 Numerical simulations with the Sugino action are currently running. 
 In this section, let us see intermediate results, in particular, ones of $\epsilon - p$, and compare them with the result
(\ref{e_minus_p}) of the gravitational theory.

 We employ the Hybrid Monte-Carlo method.  
 The integration of the fermions yields the pfaffian which is complex in general. 
 We  include the effect of  the absolute value of the pfaffian in the simulations by using the pseudo fermion method and the rational approximation of the relevant operator (see \cite{Giguere:2015cga} for details), while we simply ignore the phase effect. 
$N$ is fixed at $12$ and the two lattice sizes  $(N_t,N_x)=(8,16),(8,32)$ are employed in what follows. We take the dimensionless temperature $T_{\rm eff}=T/\lambda^{1/2}$ the values from $3$ to $0.325$, 
 and perform the simulations at thirteen different temperatures. 
In each simulation, the first $1000$ molecular dynamics (MD) time of the Hybrid Monte-Carlo method is discarded for thermalization, and we use $3000$ MD-time and choose $10$ MD-time as the bin size to evaluate $\epsilon - p$. As explained in the previous section, 
since we use the simple plaquette (\ref{P01}) to define $F_{01}$ (\ref{F01}), the extra vacua can be generated in general. However, no configurations around the extra vacua are actually observed in all simulation parameters.

The combination $\epsilon - p$ can be easily estimated from the expectation value of the action itself:
 \begin{eqnarray}
 \epsilon -p =-\frac{2}{\beta L} \langle S \rangle,
 \label{e_minus_p_in_simulatio}
 \end{eqnarray}
where $S$ is the off-shell lattice action (\ref{lat_action}) with seven auxiliary fields which are introduced to define the $Q_\pm$-transformations. 
The duality implies that $\epsilon - p$ vanishes in the zero temperature limit, and this is also realized in (\ref{e_minus_p_in_simulatio})
since supersymmetry is restored 
in the limit and the expectation value of the $Q$-exact action (\ref{lat_action}) disappears. 
 To estimate the r.h.s.~of (\ref{e_minus_p_in_simulatio}) numerically, we just evaluate the bosonic action associated with the gauge fields $U_\mu$ and the scalar fields $X_i$ 
 since the integration of auxiliary fields and fermion fields can be analytically calculated, 
 and they remain the number of degrees of freedom.

\begin{figure}[th]
\begin{center}
\hspace{-0.6cm}
\includegraphics[width=7cm,keepaspectratio,clip]{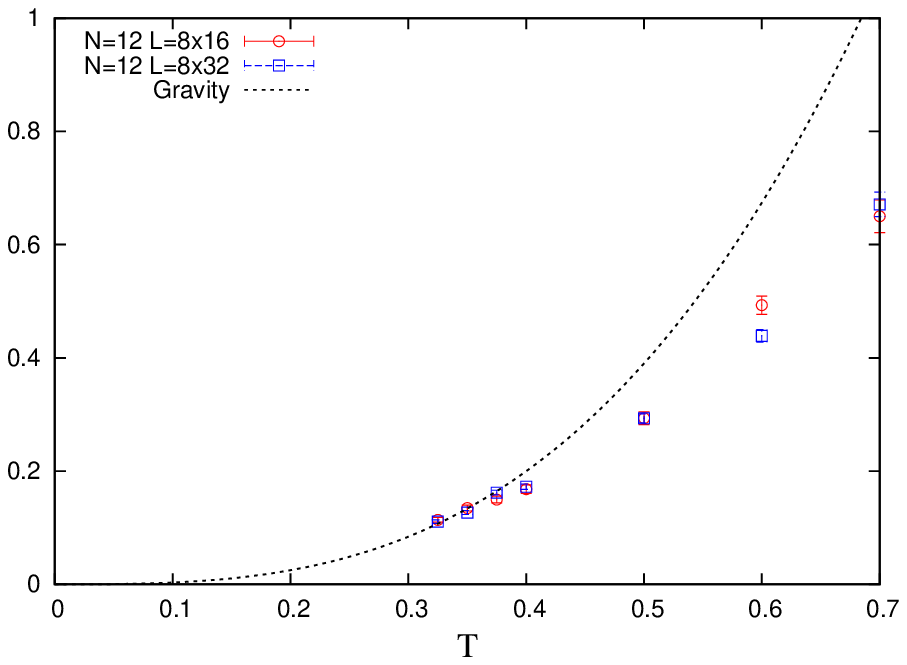}
\hspace{1cm}
\includegraphics[width=7cm,keepaspectratio,clip]{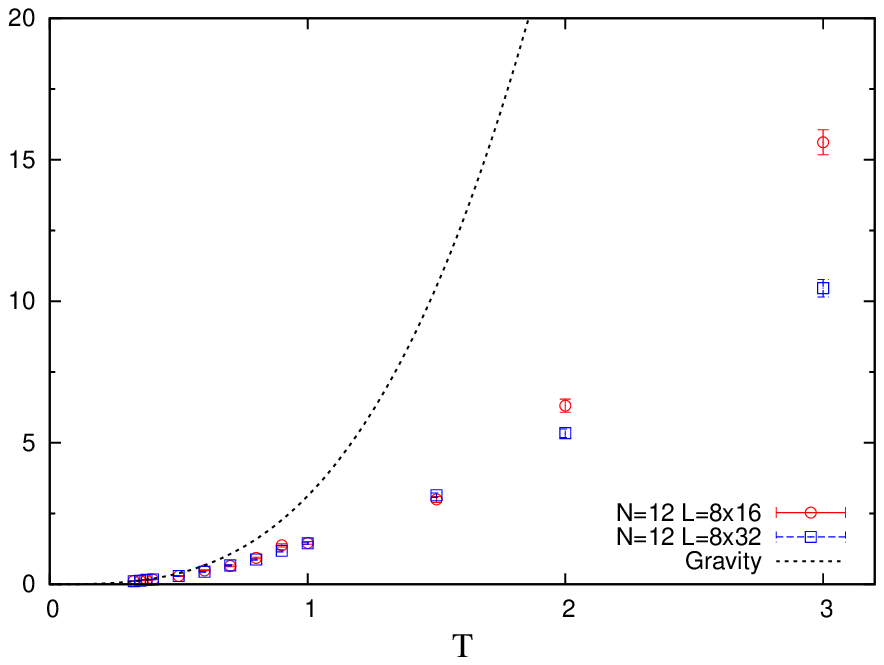}
\caption{
The energy density minus pressure normalized by $N^2$. The left figure focuses on the low temperature region of the right figure.}
\label{fig:energy_minus_pressure}
\end{center}
\end{figure}

 Fig.~\ref{fig:energy_minus_pressure} shows the results of $\epsilon - p$.  
 The $x$-axis denotes the dimensionless temperature $T_{\rm eff}$, 
 while the $y$-axis denotes the dimensionless $\epsilon - p$ normalized $N^2$.
 The red circles and  the blue squares represent the results of $(N_t,Nx)=(8,16)$ and $(8,32)$, respectively. 
 In the right figure, two results are different at high temperature, but as temperature decreases,
  they approach each other.
In the left figure focused on the low temperature region, as clearly seen, the lattice results coincide with the theoretical prediction of the gravitational theory (\ref{e_minus_p}) for $T_{\rm eff} < 0.4$.  We thus find that the duality holds in this system.

 However, we should note that this is actually the result within rather limited parameters: 
  at fixed lattice sizes and fixed $N$,  without the effect of the complex phase of the pfaffian. 
Since the deviation from the large $N$ limit starts from the term of the order of $1/N^2$ and we take $N=12$,  
our results are mostly 
saturated with respect to the large $N$ extrapolation within a few percent statistical errors. 
In addition, although the effect of the complex phase should be included in the result, it was not serious 
in one dimensional case \cite{Hanada_Nishimura}, \cite{Kadoh:2015mka}, \cite{Berkowitz:2016tyy} and 
Fig.~\ref{fig:energy_minus_pressure} could be actually considered to imply that the duality is valid and the phase effect is negligible.

 The most serious thing that changes the observation in this section is the effect of the finite lattice spacing. 
 At fixed physical temperature $T_{\rm eff}$, the (physical) lattice spacing is given by $a\lambda^{1/2}=1/(N_t T_{\rm eff})$.
 For the temperature around $T_{\rm eff}=0.4$ with $N_t=8$, the lattice spacing could not be small enough.   
 We have to take the continuum limit from further lattice simulations by using different lattice sizes to determine the final result.
 
 \section{Conclusion}
 In this paper, we have studied the duality between two dimensional  ${\cal N}=(8,8)$ SYM
and the black 1-branes solution in type IIB gravitational theory by comparing $\epsilon -p$ of both theories.
We have estimated it of the thermal SYM from the numerical simulations of the Sugino lattice action 
with two exact supersymmetries. 
The obtained  $\epsilon -p$ reproduces the theoretical prediction of the gravitational theory and implies that the duality holds in this case.

This result is obtained for $N=12$ and at fixed lattice sizes $(N_t,N_x)=(8,16) ,(8,32)$ without including the phase effect of the pfaffian.
These limitations should be taken into account when the result is quoted.
However, the large $N$ extrapolation could not be serious since $N=12$ is large enough for $\epsilon - p$ 
with a few percent statistical errors. In addition,  the effect of the phase could be much smaller than that of the lattice spacing
as was true in one dimensional case.  The effect of the finite lattice spacing becomes larger and larger as temperature decreases
with fixed $N_t$. The lattice simulations with different $N_t$ are running and 
the final result after taking the continuum limit will be presented in another paper.

The verification of  the AdS/CFT is the most important extension of this study. A few studies have already been done 
with the fine-tuning of the SUSY breaking terms \cite{Catterall:2012yq}.
As another approach without fine-tunings, it is known that the $N=4$ SYM can be constructed from the Sugino action for two-dimensional 
${\cal N}=(8,8)$ SYM via the fuzzy sphere background \cite{Hanada_Matsuura_Sugino}.
Such approach will be used to understand the AdS/CFT and further progress will be  expected in the near future.

\acknowledgments
I would like to thank Hiroshi Suzuki, Fumihiko Sugino and So Matsuura for their helpful comments on this work, 
and also thank Tsukasa Tada for valuable support. 
This work is supported by JSPS KAKENHI Grant Number JP16K05328 and the MEXT-Supported Program for the Strategic Research Foundation at Private Universities ''Topological Science'' (Grant No. S1511006).
The calculations were carried out using the RIKEN's K computer, HOKUSAI-GW, KEK's supercomputer, 
and TSC-computer of  ''Topological Science'' in Keio university. 

 \end{document}